\documentstyle[preprint,aps,epsfig]{revtex}
\tightenlines
\newcommand{\bea}{\begin{eqnarray}}
\newcommand{\eea}{\end{eqnarray}}
\newcommand{\beq}{\begin{equation}}
\newcommand{\eeq}{\end{equation}}
\newcommand{\bay}{\begin{array}}
\newcommand{\eay}{\end{array}}

\begin{document}
\preprint{\parbox{6cm}{\flushright UCSD/PTH 00-16\\[1cm]}}
\title{Bound on $\cos\alpha$ from exclusive weak  radiative $B$
decays}
\author{Dan Pirjol\\[1cm]}
\address{ Department of Physics,
University of California at San Diego, La Jolla, CA 92093}
\date{\today}
\maketitle

\begin{abstract}
We present a bound on the weak phase $\alpha$ from isospin-breaking
effects in weak radiative decays, which requires 
the CP-averaged branching ratios 
for the weak radiative decays $B^\pm\to \rho^\pm \gamma$, 
$B^0\to \rho^0/\omega \gamma$, $B\to K^* \gamma$ and the photon
energy spectrum in $B\to\gamma \ell\nu_\ell$. We carefully identify 
all sources of isospin breaking, which could possibly
mask information about the CKM parameters. They are introduced
by diagrams with photon bremsstrahlung off the spectator quark and 
diagrams with annihilation penguin topologies. The former can be 
eliminated by
combining $B\to\rho\gamma$ and $B\to K^{*}\gamma$ data, whereas
the latter effects are OZI-suppressed and can be controlled by
measuring also the $B_s\to \rho(\omega)\gamma$ modes.
The resulting bound excludes values of $\alpha$ around $90^\circ$,
provided that the combined ratio
${\cal B}(B^\pm \to \rho^\pm\gamma)/
{\cal B}(B^0 \to \rho^0/\omega\gamma)\times
{\cal B}(B^0 \to K^{*0}\gamma)/
{\cal B}(B^\pm \to K^{*\pm}\gamma)$
is found to be different from 1.
\end{abstract}

\pacs{pacs1,pacs2,pacs3}

\narrowtext

The determination of the weak phases in the Cabibbo-Kobayashi-Maskawa
(CKM) matrix is one of the main goals of physics studies at $B$
factories. The angle $\beta=-\mbox{Arg }(V_{td})$ will be measured
from the time-dependent CP asymmetry in $B\to J/\psi K_S$ decays
\cite{BaBar}.
The phase 
$\gamma = \mbox{Arg }(V_{ub}^*)$ can be determined precisely from 
time-independent measurements of rates for $B^\pm\to K^\pm D$. 
Alternatively,
it can be extracted using SU(3) symmetry from a combination of
$B^+\to K\pi$ and $B^+\to \pi^+\pi^0$ decays \cite{Grgamma}. 
On the other hand,
an extraction of the weak phase $\alpha = \pi-\beta-\gamma$ could
prove more challenging. The standard method for measuring $\alpha$
from time-dependent measurements in $B\to \pi\pi$ decays is hampered
by the need to perform a demanding isospin analysis \cite{GrLo}.

We propose in this paper a novel method for constraining the
weak phase $\alpha$ from isospin breaking effects in exclusive
radiative $B$ decays. This method makes use of the charge-averaged
rates for the weak radiative modes $B^{\pm,0}\to \rho^{\pm,0}\gamma$,
$B^{\pm,0}\to K^{*\pm,0}\gamma$ and the photon
spectrum in the radiative leptonic decay $B^\pm\to \gamma e\nu$.
All these decays are expected to have branching ratios of the order
of $10^{-6}$, which is comparable to the expected rates for the
$B\to\pi\pi$ modes needed for the traditional method for 
determining $\alpha$. 
On the positive side, the method presented here requires no tagging 
or time-dependent measurements. Also, the detection efficiency for 
these modes is very good, which is known to represent a problem with 
$B^{0}\to \pi^0\pi^0$. On the negative side, this method requires a 
binning of the photon spectrum in $B^\pm\to \gamma e\nu$, at a value 
of the photon energy equal to that in $B\to \rho\gamma$ decay
($E_\gamma = 2.6$ GeV).

Exclusive radiative $B$ decays are notorious for being plagued with
hard-to-calculate long-distance contributions, arising from
charm- and up-quark loop diagrams
\cite{VMD,LD1,LD1.5,LD2,AES,HYC,HYCeff,DoGoPe,ABS}. 
We propose here to eliminate
some of these unknown contributions, by combining
$B\to\rho(\omega)\gamma$ with $B\to K^*\gamma$ data. The remaining 
long-distance amplitudes are OZI-suppressed and hence can be argued
to be small. Their size can be controlled by measuring also the 
$B_s\to \rho(\omega)\gamma$ modes.
On the theoretical side, we rely on the fact that the dominant
long-distance amplitude, connected with the weak annihilation graph,
is exactly calculable to the leading order in an expansion in 
$1/E_\gamma$, in terms of data observable in $B\to\gamma e\nu$ decays
\cite{GrPi}.
A somewhat related method for determining $\cos\alpha$ has been 
proposed recently in \cite{GrNoRo} in terms of the exclusive 
decay $B\to\pi e^+ e^-$ on which certain cuts have been imposed. 
Unfortunately, the small rate of this mode makes this determination
very challenging from an experimental point of view.

We consider weak radiative decays of $B$ mesons into a vector meson
within SU(3) symmetry. This decay can proceed either through the
direct (short-distance) penguin transition $\bar b\to \bar s(\bar d)
\gamma$ or through the four-quark weak decay $\bar b\to \bar q_1
q_2\bar s$
(long-distance) with the photon attaching to any internal quark line.
The relevant decay amplitudes can be written as linear combinations of
a few amplitudes, each corresponding to a possible quark diagram
(see Fig.~1). The dominant amplitude is induced by the photon
penguin ${\cal Q}_7$ and is denoted with $P_t$; the gluonic penguin 
${\cal Q}_8$ is responsible for the amplitudes $G^{(i)}$, where the
index $i$ distinguishes between diagrams with the photon attaching to
the spectator quark in the $B$ meson $(i=2)$ or the remaining quark
lines $(i=1)$. The remaining amplitudes are induced by the
four-quark operators in the weak hamiltonian and consist of the
weak annihilation (WA) amplitude $A$, $W$-exchange amplitude $E$,
penguin-type amplitudes with internal $u$ and $c$ quarks
$P^{(i)}_u$, $P^{(i)}_c$ and annihilation-type penguin amplitudes 
$PA^{(i)}_u$, $PA^{(i)}_c$.
In the latter four amplitudes we distinguish again between the
spectator vs. non-spectator attachments of the photon to the quark
lines in the diagram. This decomposition in terms of graphical
amplitudes is equivalent to a more conventional SU(3) analysis 
\cite{Zepp}, and we have checked that the two approaches give the same
results.

The $B\to\rho\gamma$ decay amplitudes are given in
terms of graphical amplitudes as
(for each photon helicity $\lambda=L,R$). 
\bea\label{amp1}
A(B^-\to \rho^-\gamma_\lambda) &=&
\lambda_u^{(d)} (P^{(1)}_{u\lambda} + Q_u P^{(2)}_{uc\lambda} 
+ A_\lambda) + \lambda_c^{(d)} (P^{(1)}_{c\lambda} + Q_u
P^{(2)}_{c\lambda} )\\
& &\qquad + \lambda_t^{(d)} (P_{t\lambda} + G^{(1)}_\lambda + 
Q_u G^{(2)}_\lambda)\nonumber\\
\label{amp2}
\sqrt2 A(\bar B^0\to \rho^0\gamma_\lambda) &=&
\lambda_u^{(d)} (P^{(1)}_{u\lambda} + Q_d P^{(2)}_{u\lambda} 
 - E_\lambda - PA_{u\lambda}) +
\lambda_c^{(d)} (P^{(1)}_{c\lambda} + Q_d P^{(2)}_{c\lambda} )\\
 & &\qquad + \lambda_t^{(d)} 
(P_{t\lambda} + G^{(1)}_\lambda + Q_d G^{(2)}_\lambda +
PA_{c\lambda})\nonumber\,.
\eea
The CKM factors are defined as $\lambda_q^{(q')} = V_{qb} V_{qq'}^*$.

In the Standard Model, the amplitudes (\ref{amp1}), (\ref{amp2}) are
dominated by the short-distance penguin amplitude coupling to a
left-handed photon $P_{tL}$. The remaining long-distance amplitudes
are typically about 5-10\% of the leading $P_{tL}$ amplitude.
Therefore we will write (\ref{amp1}), (\ref{amp2}) by expanding
in the small ratios of long-distance/short-distance
contributions as
\bea
A(B^-\to \rho^-\gamma_L) &=& \lambda_t^{(d)} p_L
\left(1 +
\left|\frac{V_{ub}V_{ud}^*}{V_{tb}V_{td}^*}\right|
e^{-i(\beta+\gamma)}
\left(\varepsilon_{A}  + 
\varepsilon^{(1)}_{uc}  + \frac23 \varepsilon^{(2)}_{uc}\right)
\right) \\
\sqrt2 A(\bar B^0\to \rho^0\gamma_L) &=& \lambda_t^{(d)} p_L
\left(1 -
\varepsilon_{\rm sp}  + \varepsilon_{PA_c} +
\right.\\
& &\left.
 \left|\frac{V_{ub}V_{ud}^*}{V_{tb}V_{td}^*}\right|
e^{-i(\beta+\gamma)}
\left(-\varepsilon_{E} - \varepsilon_{PA_{uc}} +
\varepsilon^{(1)}_{uc} - \frac13
\varepsilon^{(2)}_{uc}\right)\right) \,,\nonumber
\eea
where the unitarity of the CKM matrix $\lambda_u^{(q)} + 
\lambda_c^{(q)} + \lambda_t^{(q)}=0$ has been used to eliminate
the terms proportional to $\lambda_c^{(q)}$.
We denoted here 
$p_L=P_{tL}+G_L^{(1)}+\frac23 G_L^{(2)}$ and introduced the 
small complex quantities
\bea
(\varepsilon_{\rm sp} \,,
\varepsilon_{A} \,, \varepsilon^{(i)}_{uc}\,, 
\varepsilon_{PA_c}\,, \varepsilon_{PA_{uc}})
 = \frac{1}{p_L}
(G_L^{(2)} - P_{cL}^{(2)}\,, A_L\,,
P_{uL}^{(i)}-P_{cL}^{(i)}\,,
PA_{cL}\,, PA_{uL}-PA_{cL})\,.
\eea

Let us consider the isospin-violating ratio of
CP-averaged decay rates
\bea
R_1 \equiv \frac{\tau_{B^0}}{\tau_{B^\pm}}
\frac{{\cal B}(B^\pm \to \rho^\pm\gamma)}
{2{\cal B}(B^0 \to \rho^0\gamma)}\,.
\eea
In the absence of the long-distance contributions, the amplitudes
appearing in this ratio are dominated by the penguin $P_t$, and the
value of the ratio is unity.
The deviation of $R_1$ from 1 is due to interference between
the long-distance and short-distance amplitudes.

Expanding in the small parameters $\varepsilon_i$ and keeping only
the linear terms, the isospin-violating ratio $R_1$ is given by
\bea\label{R1}
R_1 = 1 + 2\mbox{Re}\, (\varepsilon_{\rm sp} 
- \varepsilon_{PA_c} )
- 2\left|\frac{V_{ub}V_{ud}^*}{V_{tb}V_{td}^*}\right| \cos \alpha\,
\,\mbox{Re} \left(\varepsilon_{A} +
\varepsilon_{E} + \varepsilon_{PA_{uc}}
+\varepsilon^{(2)}_{uc} \right) + {\cal O}(\varepsilon_i^2)\,.
\eea
In principle there are also long-distance amplitudes contributing
to decays into right-handed photons. However they enter the ratio
$R_1$ only quadratically such that they were not written explicitly.

The long-distance amplitudes appearing in (\ref{R1}) have been 
estimated using a variety of approaches
\cite{VMD,LD1,LD1.5,LD2,AES,HYC,HYCeff,DoGoPe,ABS,GSW,EIM,pQCD,NLL}. 
Although the detailed numerical predictions differ somewhat, a 
certain hierarchy of sizes can be discerned \cite{GrPi}. The dominant 
amplitude is the $WA$ amplitude coupling to a left-handed photon $A_L$. 
The $W$-exchange amplitude $E$ is both color- and charge-suppressed
relative to $A_L$ by about a factor of 10 \cite{LD1,LD1.5,LD2} so 
that it will be neglected in the following.

The remaining long-distance amplitudes in (\ref{R1}) are considerably
more difficult to calculate. The annihilation-type penguin
amplitudes $PA_{u,c}$ can be argued to be OZI-suppressed and
hence very small. The spectator-type amplitude $\varepsilon_{\rm sp}$,
arising from diagrams with insertions of the gluonic penguin
$G_L^{(2)}$ and diagrams with charm loops $P_{cL}^{(2)}$ in which the 
photon couples to the spectator quark, is not necessarily small. 
Certain contributions of this type have been computed in \cite{pQCD} 
within the SM, and they were found to be typically of the order of 5\%
of the short-distance amplitude. It is conceivable that this amplitude
could be significant. For example, 
in certain scenarios of new physics involving an enhanced gluonic
penguin \cite{Kagan}, this amplitude can be greatly enhanced 
\cite{Petrov}.

It has been proposed in the literature \cite{LD1.5,DoGoPe,EIM,AHL}
to use a measurement of the
isospin-violating ratio $R_1$ in order to extract information on the
CKM parameters $(\rho\,,\eta)$. Given the smallness of the
expected effect, it is clear than even a value as small as
Re$\,\varepsilon_{\rm sp}=0.05$ in (\ref{R1}) could introduce
a significant uncertainty of 50\% in the constraint on $\cos\alpha$
(we used here the model estimate $\varepsilon_\rho\simeq 0.12$, see 
below).

In this Letter we point out that the long-distance amplitude
$\varepsilon_{\rm sp}$ can be
eliminated with the help of data on weak radiative $B\to K^*\gamma$
decays. The corresponding decay amplitudes can be written, in the 
SU(3) limit, in terms of the same graphical amplitudes as
those appearing in (\ref{amp1}), (\ref{amp2})
\bea\label{amp3}
A(B^-\to K^{*-}\gamma_\lambda) &=&
 \lambda_t^{(s)} (P_{t\lambda} + G^{(1)}_\lambda - P_{c\lambda}^{(1)}
+ Q_u G^{(2)}_\lambda - Q_u P_{c\lambda}^{(2)})\\
\label{amp4}
\sqrt2 A(\bar B^0\to K^{*0}\gamma_\lambda) &=&
\lambda_t^{(s)} 
(P_{t\lambda} + G^{(1)}_\lambda - P_{c\lambda}^{(1)} +
Q_d G^{(2)}_\lambda - Q_d P^{(2)}_{c\lambda})\nonumber\,.
\eea
We neglected here terms proportional to $\lambda_u^{(s)}$ which are
Cabibbo suppressed. Introducing the ratio of CP-averaged branching 
ratios for the $B\to K^*\gamma$ modes one finds
\bea
R_2 \equiv \frac{\tau_{B^\pm}}{\tau_{B^0}}
\frac{{\cal B}(B^0 \to K^{*0}\gamma)}
{{\cal B}(B^\pm \to K^{*\pm}\gamma)} = 1 - 
2\mbox{Re}\, \varepsilon_{\rm sp} +
{\cal O}(\varepsilon_i^2)\,,
\eea
where we expanded again in the small parameters 
$\varepsilon_i$ and kept only the linear terms.

The CLEO Collaboration recently measured the branching ratios for 
the exclusive $B\to K^*\gamma$ modes \cite{CLEOexcl}
\bea\label{exp1}
{\cal B}(B^\pm \to K^{*\pm}\gamma) &=& (3.76^{+0.89}_{-0.83}\pm
0.28)\times 10^{-5}\\\label{exp2}
{\cal B}(B^0 \to K^{*0}\gamma) &=& (4.55^{+0.72}_{-0.68}\pm
0.34)\times 10^{-5}\,.
\eea
Adding the statistical and systematic errors in quadrature one finds 
\beq
R_2 = 1.29\pm 0.37\,,
\eeq
where we used the lifetime ratio of charged and neutral $B$
mesons $\tau(B^\pm)/\tau(B^0) = 1.066\pm 0.024$ \cite{BLEP}.
Although the error in this determination is still large, this
leaves open the possibility of a significant long-distance
amplitude $\varepsilon_{\rm sp}$. We propose therefore
to eliminate $\varepsilon_{\rm sp}$ by using
instead of $R_1$ the following combined isospin-violating ratio
\bea\label{Rrho}
R_\rho \equiv R_1 R_2 = \frac{{\cal B}(B^\pm \to \rho^\pm\gamma)}
{2{\cal B}(B^0 \to \rho^0\gamma)}\cdot
\frac{{\cal B}(B^0 \to K^{*0}\gamma)}
{{\cal B}(B^\pm \to K^{*\pm}\gamma)} = 1 - 2\mbox{Re}\, 
\varepsilon_{PA_c} 
+ 2\mbox{Re}\,\varepsilon_\rho \cos\alpha 
+ {\cal O}(\varepsilon_i^2)\,.
\eea
We will neglect in the following the contribution of the
OZI-suppressed amplitude $\varepsilon_{PA_c}$. A measure of the
validity of this approximation could be
obtained by measuring the rate for $B_s\to\rho^0\gamma$. In the SU(3)
limit the amplitude for this decay is given by
$\sqrt2 A(\bar B_s\to\rho^0\gamma) = \lambda_u^{(s)}(E+PA_{uc}) -
\lambda_t^{(s)} PA_c$. Neglecting the contribution of the CKM
suppressed term $\lambda_u^{(s)}(E+PA_{uc})$ one finds the following 
upper bound
\beq\label{bound1}
|\varepsilon_{PA_c}|^2 \leq 
\frac{2\Gamma(B_s\to\rho^0\gamma)}{\Gamma(B^\pm\to K^{*\pm}\gamma)}\,.
\eeq
The deviation of $R_\rho$ from unity is proportional to the quantity
\bea
\varepsilon_\rho \simeq -
\left|\frac{V_{ub}V_{ud}^*}{V_{tb}V_{td}^*}\right|
|\varepsilon_A|e^{i\phi_A}
\eea
where 
we neglected the other
long-distance amplitudes $\varepsilon_i$, $i=E,PA_{uc},P_{uc}^{(2)}$.
Leaving the strong phase of the WA amplitude $\phi_A$ completely
arbitrary gives an inequality on $\cos\alpha$ provided that $R_\rho
\neq 1$
\beq\label{ineq}
|\cos\alpha| \geq |R_\rho-1|/(2|\varepsilon_\rho|)\,.
\eeq
This excludes a region in $\cos\alpha$ around $\alpha=90^\circ$,
which is of interest since this value lies within the range favored 
by  present global fits of the unitarity triangle \cite{CKMfit}.

The only ingredient missing for turning this into an observable
prediction is an estimate for $\varepsilon_\rho$. We propose to use 
for this a combination of data on $B\to\gamma e\nu$ and 
$B^\pm\to K^{\pm*}\gamma$ decays.
To leading order of an expansion in the small parameter 
$1/E_\gamma$,  the dominant $WA$ amplitudes $A_\lambda$ are given 
exactly by the factorized result \cite{GrPi}
\bea\label{ALR}
A_{L,R} = -\frac{G_F}{\sqrt2}
(C_2 + \frac{C_1}{N_c}) e m_\rho f_\rho 
\left(f_B + E_\gamma (f_A(E_\gamma) \mp f_V(E_\gamma))\right)
+ {\cal O}(\frac{\Lambda^2}{E_\gamma^2})\,.
\eea
$C_{1,2}$ are Wilson coefficients appearing in the weak nonleptonic
Hamiltonian \cite{BuBuLa}.
The formfactors $f_{V,A}(E_\gamma)$ parametrize the 
$B\to\gamma e\nu$ decay, and are defined by
\bea
& &\frac{1}{\sqrt{4\pi\alpha}}
\langle\gamma(q,\epsilon_\lambda) 
|\bar q\gamma_\mu (1-\gamma_5) b| \bar B(v)\rangle =\\
& &\qquad i\varepsilon(\mu,\epsilon_\lambda^*,v,q) 
f_V(E_\gamma) +
[\epsilon_{\lambda\mu}^* (v\cdot q) - q_\mu 
(\epsilon_\lambda^*\cdot v)]f_A(E_\gamma)\,,\nonumber
\eea
where $v$ denotes the $B$ meson velocity ($p_B=m_B v$).
They are in principle measurable from the doubly differential
spectrum of the radiative leptonic decay $B^\pm\to
\gamma e\nu$. Model calculations of these form factors suggest that 
the left-handed amplitude $A_{L}$ in $B^-$ decay is about 30\% of the
short-distance penguin amplitude $P_{tL}$ 
\cite{LD1,LD1.5,DoGoPe,GrPi} and is enhanced by about 
a factor of 7 relative to the right-handed amplitude $A_R$.
A certain simplification is obtained by considering
the form factors $f_{V,A}(E_\gamma)$ in an expansion in powers of 
$1/E_\gamma$. The leading terms in this expansion are related as
\cite{KPY}
\bea\label{f}
f(E_\gamma) \equiv 
f_V^{(B^\pm)}(E_\gamma) = \pm f_A^{(B^\pm)}(E_\gamma) +
{\cal O}(\frac{\Lambda^2}{E_\gamma^2})\,.
\eea
This implies that, to leading twist, the combination
$|V_{ub}|\cdot |f(E_\gamma)|$ can be extracted from a
measurement of the (more accessible experimentally) photon energy 
spectrum in $B^\pm\to \gamma\ell\nu_\ell$ decay.
Combining this with the rate for $B^\pm \to K^{*\pm}\gamma$, the
parameter $|\varepsilon_\rho|$ can be determined up to corrections
of order $\Lambda/E_\gamma\simeq 13\%$.
\bea\label{epsrho}
|\varepsilon_\rho|^2 = 
\frac{\displaystyle \frac{d}{dE_\gamma}\Gamma(B^\pm\to\gamma e\nu)
|_{E_\gamma = 2.6\, GeV}}{\Gamma(B^\pm\to K^{^{*\pm}\gamma})}\cdot
\frac{3(2\pi)^2(C_2+\frac{C_1}{3})^2 m_\rho^2 f_\rho^2}
{m_B^2 (m_B-2E_\gamma)}
\left(1-\frac{f_B}{2E_\gamma f(E_\gamma)}\right)^2\,.
\eea
Model calculations of the 
formfactors $f_{V,A}(E_\gamma)$ suggest that $|\varepsilon_A| \simeq
0.3$ \cite{LD1.5,DoGoPe,EHS}, which gives $|\varepsilon_\rho|\simeq 
0.12$ 
(we used here $|V_{ub}|/|V_{td}|\simeq 0.4$ \cite{AliLo}). 
The last factor in (\ref{epsrho}) is a 7\% correction and its
calculation requires theoretical input about the form factors.
In addition, the result (\ref{ALR}) implies that, to leading twist, the strong
phase $\phi_A$ vanishes. This turns the bound in (\ref{ineq})
into an identity, and the bound on $|\cos\alpha|$ into a determination
of this parameter. Alternatively, the strong phase $\phi_A$ can be
completely eliminated by combining $R_\rho$ with a measurement of the
CP asymmetry 
\beq
A_{CP}^{(\rho)}=\frac{\Gamma(B^-\to\rho^-\gamma) -
\Gamma(B^+\to\rho^+\gamma)}{\Gamma(B^-\to\rho^-\gamma) +
\Gamma(B^+\to\rho^+\gamma)} = 2|\varepsilon_\rho |\sin\alpha\sin\phi_A
 + {\cal O}(\varepsilon_i^2)\,.
\eeq

Similar results are obtained for the combined ratio involving
the decay $B\to\omega\gamma$.
The $B\to\omega\gamma$ amplitude depends on additional graphical
amplitudes $S_{u,c}$ connected with decays into an SU(3)
singlet. They are similar to the amplitudes $PA_{u,c}$ shown in
Fig.~1(d), except that the photon attaches to the other quark lines
in the diagram.
Assuming ideal mixing in the $(\omega, \phi)$ system, one has
\bea
\sqrt2 A(\bar B^0\to \omega\gamma_\lambda) &=& 
-\lambda_u^{(d)} (P_{u\lambda}^{(1)}+Q_d P_{u\lambda}^{(2)} + 
E_\lambda +\frac13 PA_{u\lambda} + \frac23 S_{u\lambda})\\
&-&
\lambda_c^{(d)}(P_{c\lambda}^{(1)}+Q_d P_{c\lambda}^{(2)}
+\frac13 PA_{c\lambda} + \frac23 S_{c\lambda}) 
-\lambda_t^{(d)}(P_{t\lambda}^{(1)} +
G_{\lambda}^{(1)}+Q_d G_{\lambda}^{(2)})\nonumber\,.
\eea
Defining again a combined ratio of decay rates as in (\ref{Rrho})
one finds to linear order in the small parameters $\varepsilon_i$
\bea\label{Romega}
R_\omega \equiv \frac{{\cal B}(B^\pm \to \rho^\pm\gamma)}
{2{\cal B}(B^0 \to \omega\gamma)}\cdot
\frac{{\cal B}(B^0 \to K^{*0}\gamma)}
{{\cal B}(B^\pm \to K^{*\pm}\gamma)} = 1 -
2\mbox{Re}\,(\frac13\varepsilon_{PA_c}+\frac23\varepsilon_{S_c}) 
+ 2\mbox{Re}\,\varepsilon_\omega \cos\alpha + {\cal O}(\varepsilon_i^2)\,.
\eea
We denoted here $\varepsilon_\omega =
-|V_{ub}V_{ud}^*|/|V_{tb}V_{td}^*|
(\varepsilon_A-\varepsilon_E+\varepsilon_{uc}^{(2)}-\frac13
\varepsilon_{PA_{uc}} - \frac23\varepsilon_{S_{uc}})$.
Keeping again only the dominant $WA$ long-distance amplitude
$\varepsilon_A$ gives that the two ratios (\ref{Rrho}), 
(\ref{Romega}) are equally sensitive to $\cos\alpha$:
$\varepsilon_\omega\simeq \varepsilon_\rho$.

Using SU(3) symmetry one can give an upper bound on the combination
of unknown long-distance amplitudes in (\ref{Romega}), similar to 
(\ref{bound1}). Neglecting a Cabibbo-suppressed term proportional
to $\lambda_u^{(s)}$, the decay amplitude for $\bar B_s\to\omega
\gamma$ is given by $\sqrt2 A(\bar B_s\to\omega\gamma) = -
\lambda_t^{(s)}(\frac13 PA_c + \frac23 S_c)$, which gives the 
inequality
\beq\label{bound2}
|\frac13\varepsilon_{PA_c}+\frac23\varepsilon_{S_c}|^2\leq
\frac{2\Gamma(B_s\to\omega\gamma)}{\Gamma(B^\pm\to K^{*\pm}\gamma)}\,.
\eeq

We turn now to the theoretical uncertainties of this method.
The neglected quadratic terms in (\ref{Rrho}) and (\ref{Romega})
can be expected to be of the order of 1-2\%. A similar
contribution is expected from the OZI-suppressed terms
proportional to $\varepsilon_{PA_c},
\varepsilon_{S_c}$ which however could be enhanced by 
rescattering effects. As explained above, the bounds (\ref{bound1}), 
(\ref{bound2}) can be used to control the size of these effects.
The SU(3) breaking effects introduce an additional uncertainty
as an incomplete cancellation of the spectator amplitude
$\varepsilon_{\rm sp}$ in $R_\rho$ and $R_\omega$. However the
corresponding effect is linear in both SU(3)
breaking and $\varepsilon_{\rm sp}$ and hence very small.

The most important theoretical limitation of this method is 
probably connected
with our ability to compute the long-distance parameters 
$\varepsilon_{\rho, \omega}$ in (\ref{Rrho}), (\ref{Romega}). 
It is unlikely that this parameter can be computed to better
than 15\% accuracy, where the uncertainty comes from higher twist
effects in (\ref{ALR}) and from the neglected long-distance
amplitudes $\varepsilon_i, i=E,PA_{uc},P_{uc}$.

However, at least in the near future, the statistical errors are 
likely to dominate the precision of any constraints on $\alpha$ 
which could be obtained from this method.
Some improvement in statistics can be achieved by noting that the
two ratios $R_\rho$ and $R_\omega$ are equal to a first
approximation. Therefore we introduce the combined ratio
\beq
R_{\rho/\omega} \equiv \frac{{\cal B}(B^\pm \to \rho^\pm\gamma)}
{{\cal B}(B^0 \to \rho^0/\omega\gamma)}\cdot
\frac{{\cal B}(B^0 \to K^{*0}\gamma)}
{{\cal B}(B^\pm \to K^{*\pm}\gamma)} \simeq 1 
+ 2|\varepsilon_\rho|\cos\phi_A \cos\alpha + {\cal O}(\varepsilon_i^2)
\eeq
for which all the conclusions derived above for $R_\rho$ and
$R_\omega$ hold unchanged.
For example, a sample of $6\times 10^8$ $B\bar B$ pairs 
(the equivalent of 500 fb$^{-1}$ integrated luminosity) would allow 
the measurement of this ratio to about 20\% accuracy, which would
give a determination of $\cos\alpha$ at a $\sigma(\langle \cos
\alpha\rangle) \simeq 24\%$ level. Even a much less precise constraint
could be useful in eliminating discrete ambiguities on $\alpha$ arising
from the method based on time-dependent measurements.
We assumed in this estimate typical branching ratios of the
order ${\cal B}(B^\pm\to\rho^+\gamma) = 1.3\times 10^{-6}$,
${\cal B}(B^0\to\rho^0\gamma) = {\cal B}(B^0\to\omega\gamma)
= 0.7\times 10^{-6}$ \cite{BaBar}, neglected the possible
backgrounds and assumed the same 
reconstruction efficiency for $\rho,\omega$ as in the CLEO 
experiment \cite{CLEOexcl}. 
Although difficult, such a measurement
could be performed in the phase I of the existing PEP-II facility,
which is expected to have collected about $8\times 10^8$ $B\bar B$
pairs by the end of 2008, or at a hadronic $B$ factory.

In conclusion, we presented a constraint on the weak phase $\alpha$
from isospin-breaking effects in exclusive weak radiative decays of 
$B$ mesons. This is similar to methods proposed earlier in
\cite{LD1.5,DoGoPe,EIM,GrNoRo,AHL} to constrain certain combinations 
of CKM parameters using similar data. We improve on these methods
by including all the relevant isospin-violating long-distance 
amplitudes. Some of these unknown long-distance amplitudes
can be eliminated by combining $B\to\rho(\omega)\gamma$ with 
$B\to K^*\gamma$ data, and the remaining ones are OZI-suppressed
and can be controlled with the help of additional $B_s$ weak 
radiative decays.

\acknowledgements

I am grateful to Ahmed Ali, Benjamin Grinstein, Alex Kagan and 
Soeren Prell for 
very useful discussions. The research of D. P. is supported in part by 
the DOE and by a National Science Foundation Grant no. PHY-9457911.

\begin{figure}[hhh]
 \begin{center}
 \mbox{\epsfig{file=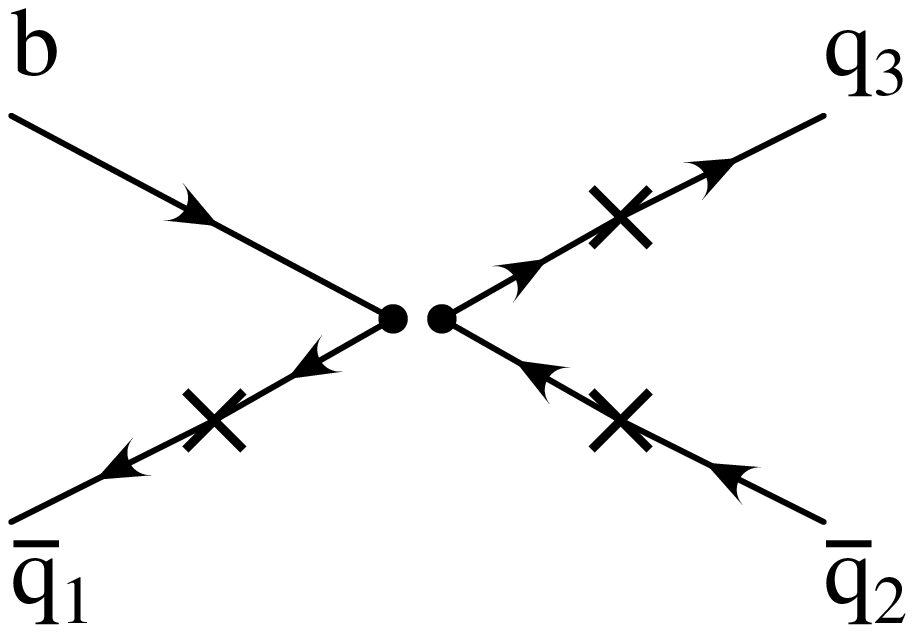,width=4cm}\qquad\qquad
 \epsfig{file=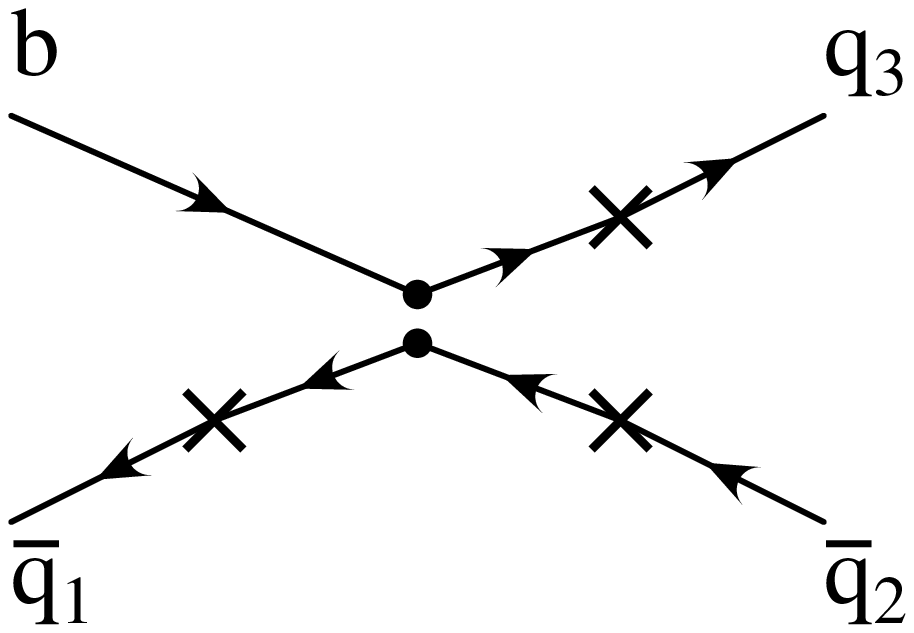,width=4cm}}\\
(a)\hspace*{5cm} (b)\\
 \mbox{\epsfig{file=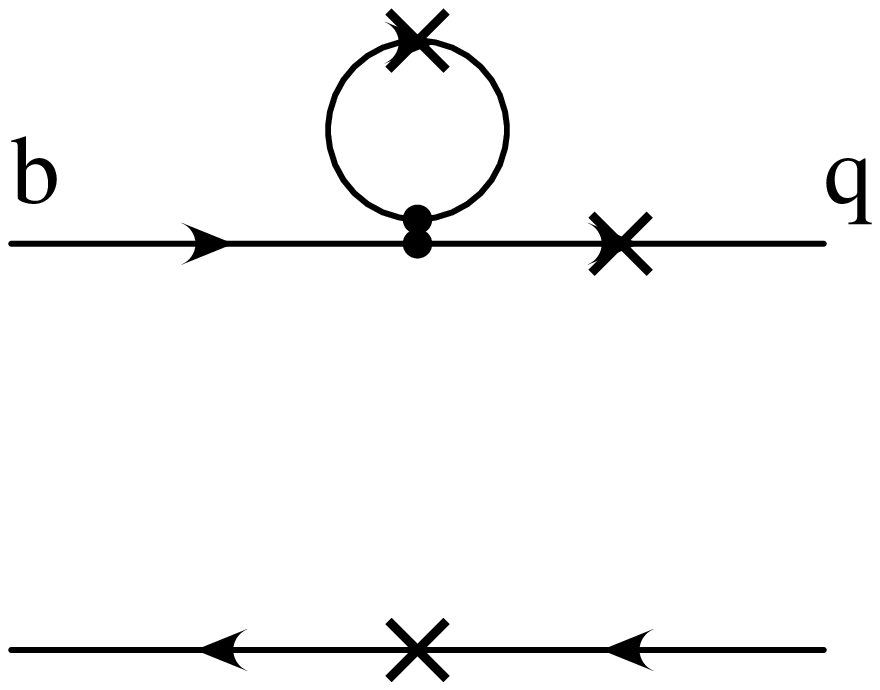,width=4cm}\qquad\qquad
 \epsfig{file=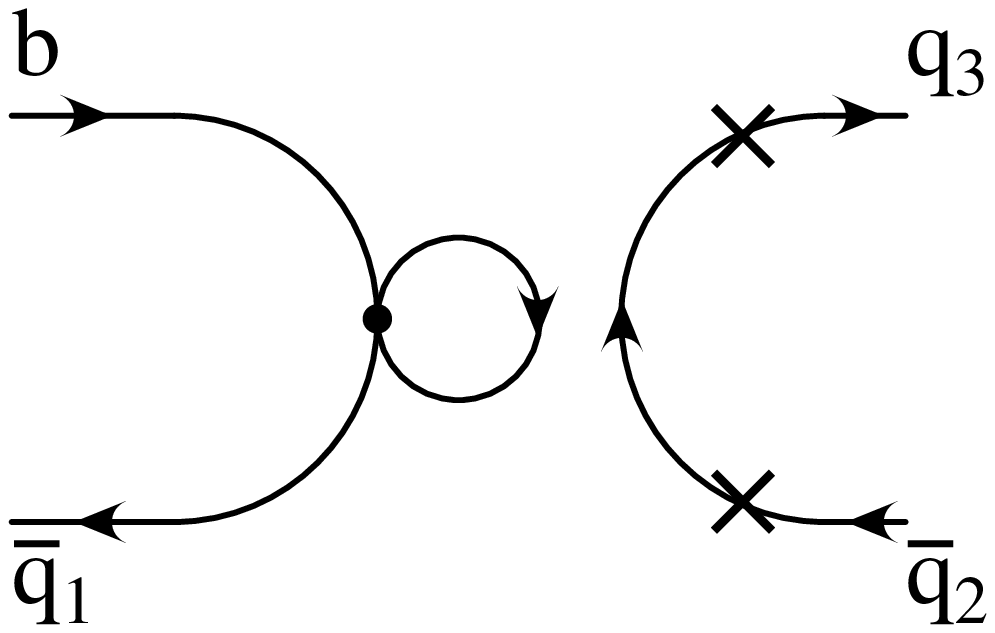,width=4cm}}\\
(c)\hspace*{5cm} (d)\\
 \mbox{\epsfig{file=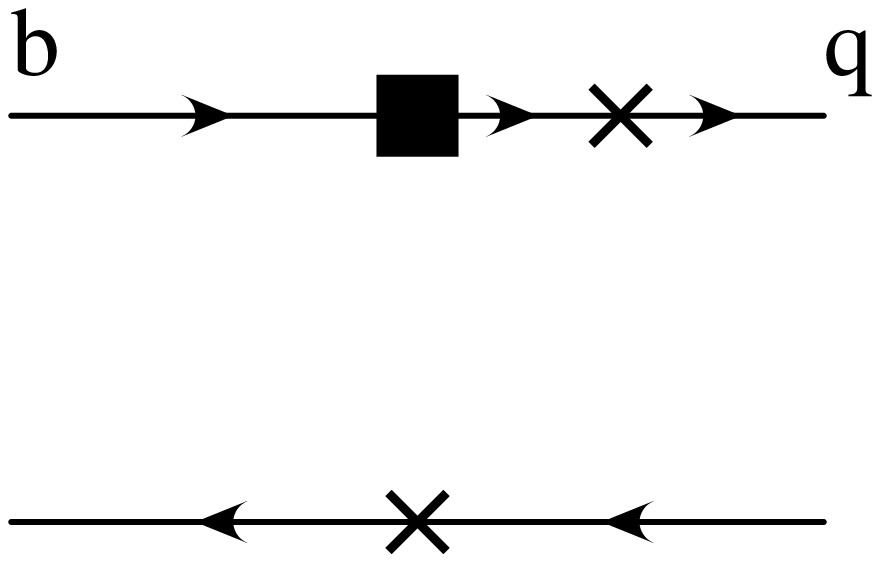,width=4cm}}\\
(e)
 \end{center}
 \caption{
Quark diagrams contributing to $\bar B\to V\gamma$ decays. The cross
marks the attachment of the photon line. a) weak annihilation
amplitude $A$; 
b) $W$-exchange amplitude $E$; c) penguin amplitudes
$P_{q'}^{(1)}$ (the photon is attached to the $\bar q=\bar d,\bar s$ 
quark or the quark $q'=u,c$ running in the loop) and $P_{q'}^{(2)}$ 
with the photon attached to the spectator quark; d) annihilation 
penguin amplitudes $PA_{q'}$; e) amplitudes with one insertion of the
gluonic penguin $G^{(1)}$ (photon attaching to the $\bar q$ line) and
$G^{(2)}$ (photon attaches to the spectator quark). The box denotes 
one insertion of the gluonic penguin operator ${\cal Q}_8$.}
\label{fig1}
\end{figure}

\end{document}